\documentclass[a4paper,11pt]{article}
\usepackage{amsmath}
\begin{document}
\title{On measuring the characteristics of quark and gluon jets in hadron-hadron collisions}
\setcounter{footnote}{0}
\author{S.G.\,Shulha\footnote{E-mail: shulga@jinr.ru}   \\ 
 Joint Institute for Nuclear Research, Dubna, Russia \\
 F. Skorina Gomel State University, Gomel, Republic of Belarus
}
\maketitle

\begin{abstract}
  To measure the characteristics of quark and gluon jets in hadron-hadron collisions, two samples of jets are used.
  Given the large statistics of jets at the LHC, the two-sample method requires taking into account the following corrections:
(1) use of measured fractions of quark and gluon jets instead of the model ones, 
(2) amendment for the contribution of jets with an unidentified flavour, 
(3) taking into account the dependence of the distributions of quark and gluon jets 
on the conditions for jet sample selections. 
  The paper presents an improved two-sample method that takes into account these corrections. \
\end{abstract}

\section{Introduction}

  The properties of hadron jets carry information about the flavour of the parent parton 
(quark or gluon), the process of parton branching and hadronization of final soft partons.
  The difference in the properties of quark and gluon jets arises from the difference in color charges:
gluon initiated jets have a larger QCD color factor leading to shower evolution with 
a higher branching probability~\cite{REVPARTPHYS:2022, INTRO-2}.

  As the properties of the $g$ and $q$ jets are different, 
then the measurement of the jet haracteristics\footnote{Short name "jet characteristic" 
    combines here the concepts of "normalized jet distribution over the jet macroparameter" 
    or raw moment of such distribution. 
	In the model, the jet can have microparameters that are not directly measured in the detector, 
    for example, the jet flavour or the parameters of the corresponding parton jet. }
in the sample, regardless of jet flavour, makes sense only 
as a first step towards finding the characteristics of $q/g$ jets.

  The jet sample consists of subsamples of $g$ jets, $q$ jets 
and jets with unidentified flavor ($x$ jets) with fractions 
$\alpha^g$, $\alpha^q$ and $\alpha^x = 1 - \alpha^g - \alpha^q$ respectively.
  The simplest characteristic of $q/g$ jets - 
the average charged-particle multiplicity inside jet - 
was measured by the CDF~\cite{CDF:2005} and ATLAS~\cite{ATLAS:2016} collaborations.
  These works used two jet samples with significantly different $\alpha^g$, determined using Monte Carlo generators.
  The contribution of $x$ jets was not taken into account.

  The measured values of $\alpha^g$ may differ from the model ones.
  An indication for this is given by the measurement of $q/g$ templates (distributions of $q/g$ jets)
for the likelihood $q/g$ discriminator, obtained in the CMS collaboration~\cite{QGL:2013, QGL:2017}.
  The ratio of measured $q/g$ templates to model ones
is a bin-by-bin factor (“Scale Factor”, SF).
  The aforementioned CMS papers used the same CDF methodology to calculate SF:
the two jet sample method with model $\alpha^g$.
  It was shown that SF for $g$ templates deviates from 1 by 25-100\%  in the region of small jet transverse momentum. 
  It was shown in~\cite{YAF-SHULHA:2024} that if calculation of SF uses the measured $\alpha^g$,
then SF is approximately equal to 1  within  uncertainty of $\alpha^g$ measurement.
  This means that the reason for the large deviation of SF from 1 
is that $g$ fractions in the data differs significantly from the generator ones~\cite{PEPAN:2021, YAF-SHULHA:2024}.

  The fraction of $g$ jets can be measured~\cite{PEPAN:2021}.
  To do this, in a certain Monte Carlo model, $q/g$ templates are created.
  Larger difference between the $q$ and $g$ templates results in more accurate $\alpha^g$ measurement.
  To enhance the differences between $q/g$ templates, $q/g$ discriminators are created~\cite{QGL:2013, QGL:2017}
which are combined jet macroparameters, built on the basis of simple jet macroparameters.
  However, $\alpha^g$ measurements can also be performed using the simple jet macroparameters~\cite{YAF-SHULHA:2024}.

  Given the large statistics of jets at the LHC, it is necessary take into account 
the admixtures of $x$ jets.
  To avoid increasing the number of fitting parameters, to measure the fraction of pure $g$ jets, 
$x$ jets can be attached to $q$ jets.
  Similarly, to measure the fractions of pure $q$ jets, generator templates for $q$ and $g+x$ jets are considered. 
  This technique allows us to obtain $\alpha^g$ and $\alpha^q$ (as well as $\alpha^x = 1 - \alpha^q - \alpha^g$) 
using two one-parameter fitting procedures~\cite{dimuonPEPAN:2024}.

  The similar way can be used to obtain the characteristics of pure $q/g$ jets 
(without admixture of $x$ jets) without increasing the number of jet samples and the number of equations.
  To find the characteristics of $q$ jets, it is necessary to use the fraction 
of pure $q$ jets, $\alpha^q$, and take $g$ fraction equal to $\alpha^g+\alpha^x \equiv 1 - \alpha^q$.
  And to find the characteristics of pure $g$ jets, it is necessary to use the fraction 
of pure $g$ jets, $\alpha^g$, and as the fraction of $q$ jets take the value $\alpha^q+\alpha^x \equiv 1 - \alpha^g$.
  Each of the two calculations uses two samples of reconstructed jets, 
which are decomposed into two subsamples of $q/g$ jets, one of which contains pure $q$ or $g$ jets, 
and the second subsample contains $q+x$ or $g+x$ jets, respectively.
  Without loss of generality, we further consider the decomposition of the jet sample into two subsamples, 
$g$ jets and $q$ jets, assuming that $x$ jets are attached to either $q$ jets or $g$ jets.
  This allows one to use two jet samples to measure the characteristics of pure $q/g$ jets, 
repeating the calculation twice.

  The two jet sample method~\cite{CDF:2005} can be used to find the $q/g$ characteristics
if the jet samples are created under identical conditions: 
identical jet kinematic regions and equal numbers of jets per event.
  The requirement for kinematic identity of jet samples means that 
the intervals of jet transverse momenta ($p_T^{jet}$) and jet pseudo-rapidity ($\eta^{jet}$) 
in jet samples are the same.
  As a rule, the dependence of the characteristics of $q/g$ jets on $p_T^{jet}$ is studied 
and the $p_T^{jet}$ interval is fixed. 
  Limiting the $\eta^{jet}$ interval will lead in practice to a decrease in statistics.
  Limiting the number of jets per event will further reduce the number of jets in the studied $p_T^{jet}$ bin. 
  This means that measurement will become impossible in many kinematic regions.

  To avoid restrictions on $\eta^{jet}$ and on the number of jets per event,
the work~\cite{PEPAN:2021} proposes to use modeling to take into account non-physical differences 
in $q/g$ characteristics between jet samples.
  The non-physical parts of jet characteristics can be parameterized for $q$ and $g$ jets 
and calculated using a Monte Carlo generator. 
  For this purpose, the concept of measures of non-universality of $q/g$ characteristics 
for a specific pair of jet samples is introduced, 
as the difference of $q/g$ characteristics between jet samples.

  In the work~\cite{PEPAN:2021}, expressions were obtained for the characteristics 
of $q/g$ jets averaged over two jet samples taking into account the non-zero measure 
of non-universality of $q/g$ jets. 
  In the next section, analytical expressions are obtained for calculating the $q/g$ characteristics in each jet sample.

\section{ Measurement of characteristics of quark and gluon jets }

  The jet sample characteristic, $X$, is decomposed into $q/g$ characteristics, $X^{q/g}$,
with fractions, $\alpha^g$ and $\alpha^ q = 1 - \alpha^g$, as weights:
\begin{equation}\label{eqs-1}
   X = \alpha^g X^g + (1 - \alpha^g) X^q. 
\end{equation}
  Eq.~(\ref{eqs-1}) can be used to measure $\alpha^g$~\cite{PEPAN:2021}.
  To do this, it is required to construct $q/g$ templates using a Monte Carlo generator  
and apply the one-parameter fitting of the measured characteristic $X$ by means of a linear combination 
of $q/g$ templates written in the right side of Eq.~(\ref{eqs-1}).
  On the other hand, if the $g$ fractions are measured, then the equation~(\ref{eqs-1}) 
can be used to measure the $q/g$ templates for various jet macroparameters. 
  To do this, it is necessary to write two equations of the type~(\ref{eqs-1}) for two jet samples:
\begin{equation}\label{eqs-2}
\begin{split}
   X_1 &= \alpha_1^g X^g + (1 - \alpha_1^g) X^q, \\
   X_2 &= \alpha_2^g X^g + (1 - \alpha_2^g) X^q.
\end{split}
\end{equation}
  Taking into account the non-universality of $q/g$ characteristics, 
the system of two equations for two jet samples contains four unknown quantities $X_{1,2}^{q/g}$:
\begin{equation}\label{eqs:eqs-withFNU}
\begin{split}
   X_1 &= \alpha_1^g X_1^g + (1 - \alpha_1^g) X_1^q, \\
   X_2 &= \alpha_2^g X_2^g + (1 - \alpha_2^g) X_2^q.
\end{split}
\end{equation}

  The difference $\Delta X^f \equiv X^f_2 - X^f_1$ is a measure of the non-universality 
of $f$ jets~\cite{PEPAN:2021} ($f = q$ or $g$). 
  The two quantities $\Delta X^f$ can be calculated using a Monte Carlo generator 
and added to the Eqs.(\ref{eqs:eqs-withFNU}) in the form of two additional relations.
  Together with the measure of non-universality, a “universal” (averaged) characteristic of $f$ jets
is introduced~\cite{PEPAN:2021}:
\begin{equation}\label{eqs:JFNU-universal}
\begin{split}
  \Delta X^f &\equiv \,\, X_2^f - X_1^f,  \\
   X^f       \equiv& \,\, \rho_1^f X_1^f + \rho_2^f X_2^f.
\end{split}
\end{equation}
  The following notations are used in the Eqs.~(\ref{eqs:JFNU-universal}):
\begin{equation}\label{eqs:JFNU-universal-where}
\begin{split}
  X^f &\equiv \frac{ n_1^f(y) + n_2^f(y) }{ N^f }, \,\,\, 
   N^f\equiv N_1^f + N_2^f,\,\,   N_{ch}^f\equiv \sum_y n_{ch}^f(y),\,\, ch=1,2,\\
  X_{ch}^f &\equiv \frac{n_{ch}^f(y)}{N_{ch}^f},\,\,\,\,\,
  \rho_{ch}^f \equiv \frac{N_{ch}^f}{N^f},\,\,  \rho_1^f + \rho_2^f = 1.
\end{split}
\end{equation}
  Here $y$ is a jet macroparameter, $n_{ch}^f(y)$ is the $y$ distribution of $N_{ch}^f$ jets with flavour $f$
for jet sample number $ch$. 
  The dependence of the jet characteristics on $y$ is omitted: $X^f\equiv X^f(y)$, $X_{ch}^f\equiv X_{ch}^f(y)$.

  Note that the quantities $\rho_{ch}^f$ introduced in the definition of universal 
(averaged) characteristics~(\ref{eqs:JFNU-universal}) are defined in the case of generator jet samples,
but not defined in the case of data. 
  We will see below that these quantities are intermediate and disappear in the final expressions 
that define two universal characteristics, $X^f$, and four non-universal characteristics, $X^f_{ch}$.

  From Eqs.~(\ref{eqs:JFNU-universal}) we find non-universal characteristics for two jet samples:
\begin{equation}\label{eqs:JFNU-nonuniversal}
\begin{split}
  X_1^f &= X^f -\rho_2^f \Delta X^f,  \\
  X_2^f &= X^f +\rho_1^f \Delta X^f.
\end{split}
\end{equation}

  Substituting~(\ref{eqs:JFNU-nonuniversal}) into~(\ref{eqs:eqs-withFNU}), we obtain 
a system of equations for the universal characteristics of $q/g$ jets, $X^{q/g}$~\cite{PEPAN:2021}:
\begin{equation}\label{eqs:DDR-withJFNU}
\begin{split}
   \tilde{X}_1 &= \alpha_1^g X^g + (1 - \alpha_1^g) X^q, \\
   \tilde{X}_2 &= \alpha_2^g X^g + (1 - \alpha_2^g) X^q,
\end{split}
\end{equation}
  The left sides of the Eqs.~(\ref{eqs:DDR-withJFNU}) contain the measured 
characteristics with corrections for non-universality, having the form~\cite{PEPAN:2021}:
\begin{equation}\label{eqs:QGL-FIT-withFNU-tilde-where1}
\begin{split}
 \tilde{X}_1 &\equiv X_1 + \gamma^{\rm DAT}\Delta X^{\rm JFNU},   \\
 \tilde{X}_2 &\equiv X_2 - \Delta X^{\rm JFNU}.
\end{split}
\end{equation}
  The following notations are used in Eq.~(\ref{eqs:QGL-FIT-withFNU-tilde-where1}):
\begin{equation}\label{eqs:QGL-FIT-withFNU-tilde-where2}
\begin{split}
 \Delta X^{\rm JFNU} &\equiv \beta^q \Delta X^q + \beta^g \Delta X^g, \\
 \beta^f \equiv \frac{\alpha_1^f\alpha_2^f}{\alpha_1^f + \gamma^{\rm DAT} \alpha_2^f},\,\,\, 
  \gamma^{\rm DAT} &\equiv \frac{N_2}{N_1}, \,\,\, f=q,g,\,\, \alpha_{ch}^q = 1 - \alpha_{ch}^g,
\end{split}
\end{equation}
where $N_{ch} \equiv N_{ch}^q + N_{ch}^g $ is the number of jets in jet sample with number $ch$.

  Solving the system of equations~(\ref{eqs:DDR-withJFNU}), we find 
the universal characteristics $X^{q,g}$ in terms of the quantities $\tilde{X}_{1,2}$ 
and $\alpha_{1 ,2}^g $ (see (13) in~\cite{PEPAN:2021}).
  Substituting the $X^{q,g}$ found in this way into Eqs.(\ref{eqs:JFNU-nonuniversal}), 
we obtain expressions for the nonuniversal $q/g$ characteristics in terms 
of the measured quantities $X_{1,2}$ and two measures of nonuniversality $\Delta X^{q/g}$:
\begin{equation}\label{eqs:NonUniversal}
\begin{split}
   X^q_1 &= \frac{\alpha_2^g X_1 - \alpha_1^g X_2}{\alpha_2^g - \alpha_1^g} 
   + (\Delta X^g \alpha_2^g +  \Delta X^q \alpha_2^q)\frac{\alpha_1^g}{\alpha_2^g - \alpha_1^g}, \\
&\alpha^q_{ch} \equiv   1 - \alpha^g_{ch}.
\end{split}
\end{equation}
  The remaining three quantities ($X^q_2$, $X^g_1$ and $X^g_2$) can be written by performing 
formal substitutions in Eq.~(\ref{eqs:NonUniversal}): $\{ q \leftrightarrow g \}$ and $\{ 1 \leftrightarrow 2 \}$.
  These expressions do not contain the fraction of $f$ jets from the sample $ch$ 
in the combined (1+2) jet sample introduced in the intermediate calculations 
($\rho^f_{ch}$, see~(\ref{eqs:JFNU-universal-where})) 
and do not contain the ratio of the number of jets in jet samples $\gamma^{\rm DAT}$ 
defined in~(\ref{eqs:QGL-FIT-withFNU-tilde-where2}).

\section{Conclusion}
  Using two samples of jets, it is possible to calculate the characteristics 
of $q/g$ jets (distributions of $q/g$ jets and raw moments of these distributions), 
assuming that the $g$ fractions in jet samples are known.
  The work presents an improved two-sample method, which uses 
the fractions of $q/g$ jets, previously measured, takes into account 
the dependence of the $q/g$ jet characteristics on the jet sample 
and the contribution of jets with an unidentified flavour ($x$ jets), 
the fraction of which can also be measured.
  The $q/g$ jet characteristics for each of the two jet samples 
are expressed in terms of the measured characteristics in the two jet samples,
the measured $g$ fractions, and measures of the non-universality of $q/g$ jets -
the differences in $q/g$ jet characteristics between the samples, which can be determined  by simulation.
  To take into account the contribution of $x$ jets, 
the calculation method must be applied twice for one pair of jet samples: 
to calculate the characteristics of pure $q$ jets, the fraction of pure $q$ jets 
and the fraction of $g+x$ jets are used, 
and to calculate the characteristics of pure $g$ jets, 
the fraction of pure $g$ jets and the fraction of $q+x$ jets are used.
  The resulting formulas make it possible to significantly expand the areas 
of jet selection for measuring the characteristics of $q/g$ jets.

\end{document}